\documentclass[conference]{IEEEtran}
\usepackage{cite}
\usepackage{amsmath,amssymb,amsfonts}

\usepackage{algorithmic}
\usepackage{subfigure}

\usepackage{graphicx}
\graphicspath{{./figures/}}

\newcommand{\be}{\begin{equation}}
\newcommand{\ee}{\end{equation}}

\usepackage[bookmarks=true]{hyperref}
\hypersetup
{
    colorlinks=true, 
    linktoc=all,     
    linkcolor=black, 
    allcolors=black, 
    bookmarksopen=true,
    bookmarksnumbered=true,
    pdfstartview=, 
    pdfremotestartview=, 
}
\usepackage[all]{hypcap}
\usepackage{bookmark}
\usepackage{cleveref}

\def\BibTeX{{\rm B\kern-.05em{\sc i\kern-.025em b}\kern-.08em
    T\kern-.1667em\lower.7ex\hbox{E}\kern-.125emX}}
    
\begin{document}

\title{\huge{Robust Classification of Digitally Modulated Signals \\ Using Capsule Networks and Cyclic Cumulant Features}}

\author{\IEEEauthorblockN{John A. Snoap, James A. Latshaw, and Dimitrie C. Popescu}
\IEEEauthorblockA{ECE Department, Old Dominion University \\
Norfolk, VA 23529, USA \\
\{jsnoa001, jlats001, dpopescu\}@odu.edu \vspace{-0.5cm} }
\and
\IEEEauthorblockN{Chad M. Spooner}
\IEEEauthorblockA{NorthWest Research Associates \\
Monterey, CA 93940, USA \\
cmspooner@nwra.com \vspace{-0.5cm} }
}

\maketitle

\begin{abstract}
The paper studies the problem of robust classification of digitally modulated signals using capsule networks and cyclic cumulant (CC) features extracted
by cyclostationary signal processing (CSP). Two distinct datasets that contain similar classes of digitally modulated signals but that have been generated
independently are used in the study, which reveals that capsule networks trained using CCs achieve high classification accuracy while also outperforming
other deep learning-based approaches in terms of classification accuracy as well as generalization abilities.
\end{abstract}


\section{Introduction}\label{sec:Intro}
Blind classification of digitally modulated signals is a challenging problem that occurs in both military and commercial applications such as
spectrum monitoring, signal intelligence, or electronic warfare \cite{Dobre_Sarnoff2005}. Conventional approaches to modulation classification
use signal processing techniques that include likelihood-based methods \cite{Hameed_etal_TW2009, Xu_etal_TSMC2011} or CSP-based
techniques~\cite{Spooner_Asilomar2000}. In recent years, machine learning (ML) techniques have also been explored for classifying digitally
modulated signals, and in this direction we note the work in \cite{Latshaw_COMM2022, Snoap_CCNC_2022, Tim2018, Tim2017}, which uses
the I/Q signal components for training and signal recognition/classification, or the alternative approaches in \cite{Sun2018, Zhang2020,
Kulin2018, Rajendran2018, Bu2020}, in which the amplitude/phase or frequency domain representations are used.

The ML-based approaches to blind modulation classification employ neural networks (NNs) and require extensive training to be able to distinguish
different classes of digitally modulated signals. Their classification performance depends on the characteristics of the training dataset, and the
robustness of the ML-based classifiers can be affected by factors such as under- or over-fitting the training data \cite{Bilbao_ICICIS2017} or
lack of generalization~\cite{Djolonga_etal_arXiv2020}. As noted in \cite{Latshaw_COMM2022, Snoap_CCNC_2022},  NNs using the I/Q
signal data for digital modulation classification have difficulty maintaining good performance when tested on signals coming from datasets that
were not used in the training process. This is an important aspect in wireless communication systems, where many parameters can affect the
I/Q components of the wireless signals and ensuring robustness of the ML-based NN classifier by creating an exhaustive dataset that accounts
for all possibilities is impractical to achieve.

Motivated by the lack of robustness of ML-based classifiers using I/Q signal components, we present a new approach to ML-based classification
of digitally modulated signals. This involves the use of capsule NNs, which outperform other types of NNs in the classification of digitally modulated
signals \cite{Latshaw_COMM2022}, in conjunction with CC features obtained by CSP \cite{Spooner_Asilomar1995}, which are known for their
robustness to co-channel signals~\cite{Spooner_Asilomar2001} or variations in noise models~\cite{Hazza_etal_EurasipWCN2011}. We note the
related approach in \cite{SCF_NN_2021}, which uses a convolutional NN and the spectral correlation function (SCF) in the context of spectrum
sensing to classify wireless signals based on standards such as GSM, UMTS, and LTE.

To assess the robustness and generalization abilities of the proposed ML-based digital modulation classifier we use two distinct datasets that
are publicly available from \cite{CSPblog_DataSets} and  include signals with similar digital modulation schemes but which have been generated
independently using distinct parameters, such that data from the testing dataset is not used in the training dataset. 

The remainder of this paper is organized as follows:  we provide related background information on CSP and ML for digital modulation classification
in Section~\ref{sec:Background}, followed by presentation of the proposed NNs in Section~\ref{sec:Details}. In Section~\ref{sec:Datasets} we provide
details on the datasets used for training and testing the NNs, and provide numerical performance results in Section~\ref{sec:NumericalResults}.
We conclude the paper with discussion and final remarks in Section~\ref{sec:Conclusion}.

\section{CSP and ML Background for \\ Digital Modulation Classification}\label{sec:Background}
CSP provides a set of analytical tools for estimating distinct features that are present in digital modulation schemes and that can be used
to perform classification of digitally modulated signals in various scenarios involving stationary noise and/or cochannel interference. These tools include higher-order CC estimators \cite{Spooner_Asilomar2001, Spooner_Asilomar1995} and
SCF estimators \cite{Gardner_CSP_pt1_1994, Spooner_CSP_pt2_1994} that can be compared to a set of theoretical CCs or SCFs for
classifying the digital modulation scheme embedded in a noisy signal.

ML-based classification of digitally modulated signals has received increased attention lately as a promising alternative to the conventional CSP-based
methods. In this direction, we note the ML-based approaches using convolutional NNs (CNNs) and residual NNs (RESNETs) in \cite{Snoap_CCNC_2022,
Tim2018} as well as capsule NNs \cite{Latshaw_COMM2022}, which have been shown to achieve high performance at the task of digital modulation
classification but with low robustness and generalization performance. These approaches use I/Q signal data for classification, which is affected by
the propagation channel or other effects with underlying random variables whose probability distributions widely vary in practice. To name
a few of these variables: 
\begin{itemize}
\item The symbol interval, $T_0$, and symbol rate, $1/T_0$;
\item The carrier-frequency offset (CFO), $f_0$;
\item The excess bandwidth implied by the square-root raised-cosine (SRRC) roll-off parameter, $\beta$;
\item The signal power level, which directly impacts the in-band signal-to-noise (SNR) ratio.
\end{itemize}
While some of these parameters may have limited practical ranges (such as $\beta$ for example, typically in the $[0.2, 0.5]$ range), others
possess an infinite number of valid practical choices (such as $T_0$ or $f_0$, for example).

Since ML-based approaches use NN models that learn by observing training examples and signal generation parameters range over an infinite set
of possibilities, it is necessary to ensure a trained NN can not only classify modulated signals when the signal generation parameters conform to its
training examples, but that it can also correctly identify modulation types when the signal generation parameters fall outside the distribution of the
training examples. In this direction, our proposed approach uses CC features for classification of digitally modulated signals, which are obtained
through blind estimation methods from the I/Q data using a combination of traditional and cyclostationary signal processing.

\vspace{-0.1cm}
\subsection{CC Feature Extraction}
\vspace{-0.1cm}
The CSP-based approach employed for feature extraction prior to NN training is based on \cite{Spooner_Asilomar2001}, where CC values are blindly
estimated for each example.  We note that CCs are not a form of higher-order statistics (HOS) that produce time-invariant higher-order moments and
time-invariant higher-order cumulants; but rather come from higher-order cyclostationarity (HOCS) where the signal models assume cyclostationarity
rather than stationarity and so are mathematically different from traditional HOS signal models.  HOCS involves time-varying higher-order moments
and time-varying higher-order cumulants, from which cyclic moments and CCs are obtained through a Fourier series.

Given a generic data model:
\begin{align}
x \left( t \right) = \sum_{b=0}^{M-1}{a_b s_b \left( t-t_b \right) e^{i 2 \pi f_b t + i \phi_b} + w \left( t \right)}
\end{align}
each component signal $s_b \left( \cdot \right)$ is the complex envelope of a corresponding radio frequency (RF) signal, and $w \left( t \right)$ is
AWGN. The $n$\textsuperscript{th}-order moment function is defined by
\begin{align}
R_x \left( t, \boldsymbol{{\tau}}; n, m \right) =
\widehat{E}^{ \left \lbrace \beta \right\rbrace } { \left \lbrace \prod_{j=1}^{n}{x^{ \left( \ast \right) j } \left( t + \tau_j \right) } \right \rbrace }
\end{align}
where $m$ of the factors are conjugated, $\left( \ast \right)$ represents an optional conjugation, and $\widehat{E}^{ \left\lbrace \beta \right \rbrace }$
is the sine wave extraction operator:
\begin{align}
\widehat{E}^{ \left \lbrace \beta \right \rbrace }{\left\lbrace g \left( t \right) \right \rbrace} = \sum_{\beta}{g_{\beta} e^{i 2 \pi \beta t}}
\end{align}
\begin{align}
g_{\beta} \triangleq \lim_{T \to \infty}{\dfrac{1}{T} \int_{\frac{-T}{2}}^{\frac{T}{2}}{g \left( u \right) e^{- i 2 \pi \beta u}}\,du} \equiv
\langle g \left( u \right) e^{- i 2 \pi \beta u} \rangle
\end{align}

The corresponding $n$\textsuperscript{th}-order cumulant function is given by:
\begin{align}
C_x \left( t, \boldsymbol{{\tau}}; n, m \right) = \sum_{P_n}{\left[ h \left( p \right) \prod_{j=1}^{p}{R_{x_{\nu_j}} \left( t, \tau_{\nu_j}; n_j, m_j \right) } \right] }
\end{align}
where the sum is over all distinct partitions $ \left\lbrace \nu_j \right\rbrace _{j=1}^{p} $ of the index set $ \left\lbrace 1,2,...,n \right\rbrace $ and
$h \left( p \right) = \left( -1 \right) ^{p-1} \left( p-1 \right) !$.  Since the $n$\textsuperscript{th}-order moment functions are polyperiodic functions
of time, so too are the $n$\textsuperscript{th}-order cumulant functions, and we can represent the latter as a Fourier series. The coefficients of
this Fourier series are given by
\begin{align}\label{eq:CCexpr}
C_x^{\alpha} \left( \boldsymbol{{\tau}}; n, m \right) = \langle C_x \left( t, \boldsymbol{{\tau}}; n, m \right) e^{- i 2 \pi \alpha t} \rangle
\end{align}
where $C_x^{\alpha} \left( \boldsymbol{{\tau}}; n, m \right)$ is called a cyclic cumulant (CC) and $\alpha$ is an $n$\textsuperscript{th}-order
cycle frequency (CF).

\begin{figure}
\begin{center}
{\includegraphics[width=\linewidth]{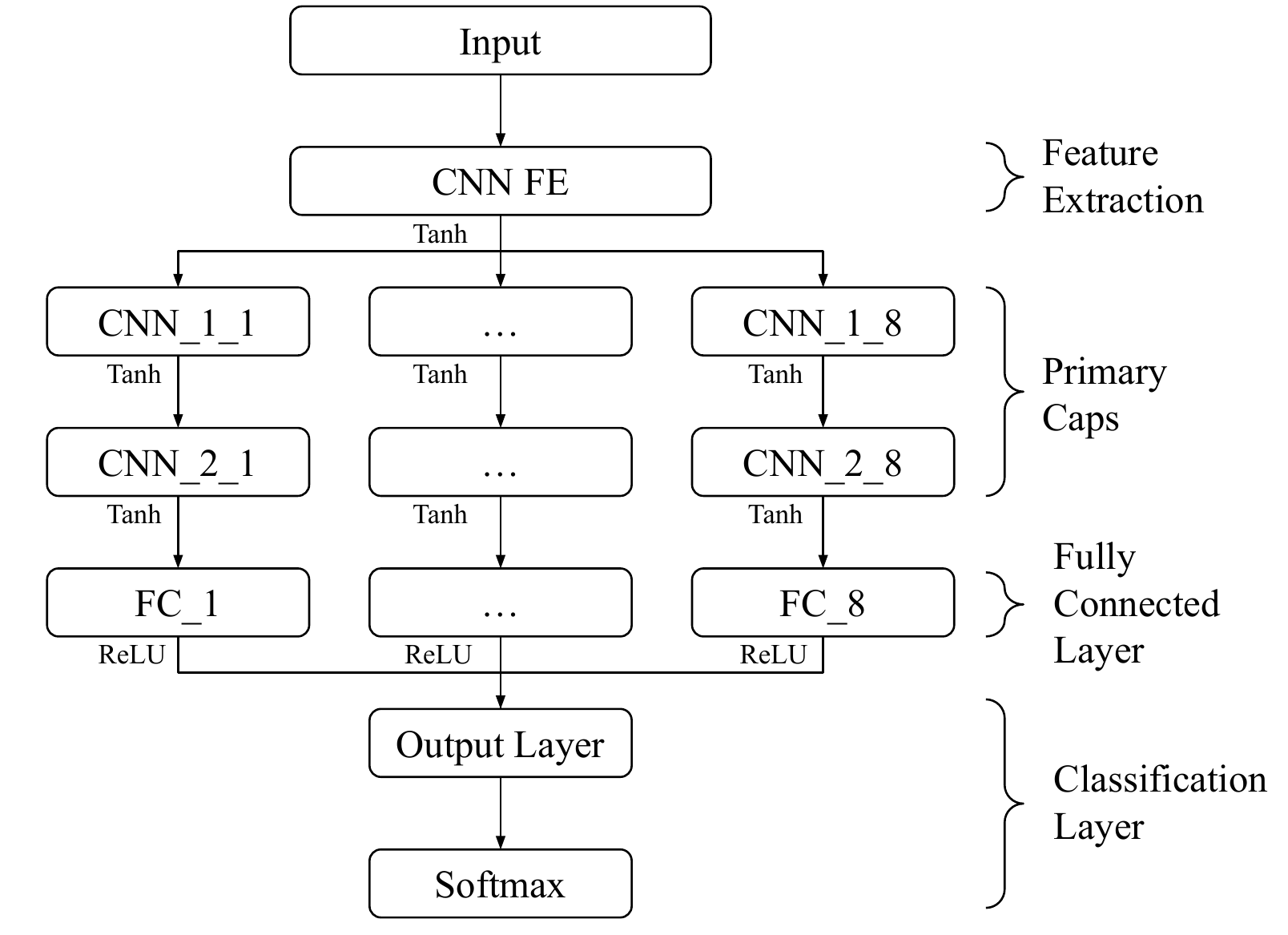}}
\caption{Capsule network topology with eight branches for classifying the eight digital modulation schemes of interest.}\label{fig:CAP_Topology}
\end{center}
\vspace{-0.5cm}
\end{figure}

\vspace{-0.1cm}
\subsection{Capsule Networks for Digital Modulation Classification}
Capsule networks perform classification based on the hierarchical features present in the input data by means of capsules which focus on activating for
class-specific features present in the input data \cite{Sabour_etal_NIPC2017}. For digital modulation classification, the capsule network will
activate on the presence of features (and not necessarily the order of occurrence of said features) in the input data which correspond to the likelihood
that a specific class is present in the input data, and in this paper, we consider the capsule network with topology illustrated in Fig.~\ref{fig:CAP_Topology},
which is similar to the one used in \cite{Latshaw_COMM2022} for classifying eight digital modulation schemes of interest:  BPSK, QPSK, 8-PSK,
$\pi/4$-DQPSK, MSK, 16-QAM, 64-QAM, and 256-QAM.  

A CNN with structure similar to the one in \cite{Snoap_CCNC_2022} will also be adapted for modulation classification using CC features and will be
trained and tested on the same data as the capsule network. The classification performance and robustness of the capsule network will be compared
to that of the CNN, as well as with that of alternative approaches where capsule NNs and CNNs use I/Q signal data for signal classification 
\cite{Latshaw_COMM2022, Snoap_CCNC_2022}.

\section{Details of the NNs and CC Feature Estimation}\label{sec:Details}
\vspace{-0.1cm}
The specific characteristics of the capsule network parameters we designed for classifying digitally modulated signals using CC
features are outlined in Table~\ref{table:CC_CAPnetwork}, while those of the  alternative CNN developed for comparison purposes
are given in Tables~\ref{table:CC_CNNnetwork}, \ref{table:ConvMaxPoolLayer}, and \ref{table:ConvAvgPoolLayer}.

\begin{table}[!tp]
\centering
\caption{CC-trained Capsule Network Layout}
{
\vspace{-0.25cm}
\begin{tabular}{ c c c c}
\hline\hline
Layer & (\# Filters)[Filt Size] & Stride & Activations \\
\hline
Input					&									&					& $11 \times 15 \times 1$ \\
Conv					& ($56$)[$6 \times 4 \times 1$]	& [$1 \times 2$]	& $11 \times 8 \times 56$ \\
Batch Norm				& \\
Tanh					& \\
Conv-1-(i)				& ($56$)[$4 \times 4 \times 56$]	& [$1 \times 2$]	& $11 \times 4 \times 56$ \\
Batch Norm-1-(i)		& \\
Tanh-1-(i)				& \\
Conv-2-(i)				& ($72$)[$4 \times 6 \times 56$]	& [$1 \times 2$]	& $11 \times 2 \times 72$ \\
Batch Norm-2-(i)		& \\
Tanh-2-(i)				& \\
FC-(i)					& &	& $7$ \\
Batch Norm-3-(i)		& \\
ReLu-1-(i)				& \\
Point FC-(i)			& & & $1$ \\
Depth Concat(i=1:8)		& & & $8$ \\
SoftMax					& \\
\hline\hline
\end{tabular}
\vspace{-0.25cm}
}\label{table:CC_CAPnetwork}
\end{table}

\begin{table}[!tp]
\centering
\caption{CC-trained CNN Layout}
{
\vspace{-0.25cm}
\begin{tabular}{ c c c }
\hline\hline
Layer & (\# Filters)[Filter Size] & Activations \\
\hline
Input			& 							& $165 \times 1$ \\
ConvMaxPool		& ($16$)[$23 \times 1$]	& $82 \times 16$ \\
ConvMaxPool		& ($24$)[$23 \times 16$]	& $41 \times 24$ \\
ConvMaxPool		& ($32$)[$23 \times 24$]	& $20 \times 32$ \\
ConvMaxPool		& ($48$)[$23 \times 32$]	& $10 \times 48$ \\
ConvMaxPool		& ($64$)[$23 \times 48$]	& $5 \times 64$ \\
ConvAvgPool		& ($96$)[$23 \times 64$]	& $96$ \\
FC/SoftMax		& 				& \# Classes \\
\hline\hline
\end{tabular}
\vspace{-0.25cm}
}\label{table:CC_CNNnetwork}
\end{table}

\begin{table}[!tp]
\centering
\caption{ConvMaxPool Layer}
{
\vspace{-0.25cm}
\begin{tabular}{ c c c c }
\hline\hline
Layer 		& (\# Filters)[Filt Size] & Stride & Activations \\
\hline
Input		& ($A$)[$B \times C$]	& 					& $X \times Y$ \\
Conv		& ($A$)[$B \times C$]	& [$1 \times 1$] 	& $X \times (Y \cdot A/C)$ \\
Batch Norm	& 						& 					& $X \times (Y \cdot A/C)$ \\
ReLU		& 						& 					& $X \times (Y \cdot A/C)$ \\
Max Pool	& ($1$)[$1 \times 2$]	& [$1 \times 2$]	& $(X/2) \times (Y \cdot A/C)$ \\
\hline\hline
\end{tabular}
\vspace{-0.25cm}
}\label{table:ConvMaxPoolLayer}
\end{table}

\begin{table}[!tp]
\centering
\caption{ConvAvgPool Layer}
{
\vspace{-0.25cm}
\begin{tabular}{ c c c c }
\hline\hline
Layer 		& (\# Filters)[Filter Size] & Stride & Activations \\
\hline
Input		& ($A$)[$B \times C$]	& 					& $X \times Y$ \\
Conv		& ($A$)[$B \times C$]	& [$1 \times 1$] 	& $X \times (Y \cdot A/C)$ \\
Batch Norm	& 						& 					& $X \times (Y \cdot A/C)$ \\
ReLU		& 						& 					& $X \times (Y \cdot A/C)$ \\
Avg Pool	& ($1$)[$1 \times X$]	& [$1 \times 1$]	& $1 \times (Y \cdot A/C)$ \\
\hline\hline
\end{tabular}
\vspace{-0.4cm}
}\label{table:ConvAvgPoolLayer}
\end{table}

To obtain accurate estimates of the CC features used for training and signal classification, knowledge of signal parameters such as the symbol rate
and CFO are necessary because they define the cycle frequencies needed for CC computations. These signal parameters can be obtained using
second-order cyclostationarity techniques that estimate the SCF such as the strip spectral correlation analyzer (SSCA)~\cite{Brown_SSCA_1993},
or the time- and frequency-smoothing methods \cite{Gardner_CSP_pt1_1994, Spooner_CSP_pt2_1994}. Additional parameter estimates such as
the excess bandwidth and in-band SNR further refine the CC estimates and can be obtained using band-of-interest detectors that do not require
any CSP \cite{BOIdetector}.

To reduce computations for our estimates of the CC (\ref{eq:CCexpr}) we use the following parameters:
\begin{itemize}
\item The delay vector $\boldsymbol{{\tau}} = 0$;
\item The orders of CC features are limited to $n = \left \lbrace 2, 4, 6 \right \rbrace$\footnote{The number of conjugation choices is constrained
by the order to $n+1$.}.
\item For each $\left( n, m \right)$ pair, the CFs where CCs are non-zero are related to the CFO ($f_0$) and symbol rate ($1/T_0$) by
\be
\alpha = (n-2m)f_0 \pm k/T_0,
\ee
where $k$ is the set of non-negative integers restrained to a maximum value of $k=5$.
\end{itemize}
These settings imply a total of $11$ CFs for each of the $15$ $\left( n, m \right)$ pairs, or 165 CC estimates for each signal example, matching
the dimension of the input layers for NNs considered.

We note that higher-order CCs generally have larger magnitudes than lower-order CCs and that the CCs also scale with signal power. Because of
these effects, the CC estimates $\widehat{C}_x^{\alpha} \left( \boldsymbol{{\tau}}; n, m \right)$ are further processed prior to use with NN training, as follows:
\begin{itemize}
\item The CC estimates are first warped \cite{Spooner_Asilomar2000,Spooner_Asilomar2001}  by their order using
$\widehat{C}_x^{\alpha} \left( \boldsymbol{{\tau}}; n, m \right)^{ \left( 2/n \right) }$. 
\item The warped CC estimates are subsequently scaled to a signal power of unity, using a blind estimate of the signal power\footnote{This provides
consistent values for the neural network to train on and prevents varying signal powers from causing erroneous classification results through saturation
of neurons -- a common issue with input data that does not go through some normalization process.}.
\end{itemize}
After the additional processing, the capsule network and the CNN can be trained using the warped and scaled CC estimates.

\vspace{-0.2cm}
\section{Datasets Used for NN Training and Testing}\label{sec:Datasets}
To both design and assess the robustness and generalization abilities of the proposed NNs, we use two publicly available datasets that both contain signals
corresponding to the eight digital modulation schemes of interest (BPSK, QPSK, 8-PSK, $\pi/4$-DQPSK, MSK, 16-QAM, 64-QAM, and 256-QAM). The two
datasets are available from \cite{CSPblog_DataSets} as \texttt{CSPB.ML.2018} and \texttt{CSPB.ML.2022}, respectively, and the signal generation parameters
for both datasets are listed in Table~\ref{table:SigGenParms}.
\vspace{-0.25cm}
\begin{table}[!htbp]
\centering
\caption{Dataset Signal Generation Parameters}
{
\vspace{-0.25cm}
\begin{tabular}{ c c c }
\hline\hline
Parameter & \texttt{CSPB.ML.2018} & \texttt{CSPB.ML.2022} \\
\hline
Sampling Frequency, $f_s$ & 1 Hz & 1 Hz \\
CFO, $f_0$ & $U(-0.001, 0.001)$  & Omitted \\
Symbol Period, $T_0$, Range & $[1, 23]$ & $[1, 29]$ \\
Excess Bandwidth, $\beta$, Range & $[0.1, 1]$ & $[0.1, 1]$ \\
In-Band SNR Range (dB) & $[0, 12]$ & $[1, 18]$ \\
SNR Center of Mass & $9$ dB & $12$ dB \\
\hline\hline
\end{tabular}
}\label{table:SigGenParms}
\end{table}

The omission of the \texttt{CSPB.ML.2022} CFO is intentional because of how easy it would be to generate new training samples
by frequency shifting signals in \texttt{CSPB.ML.2018} to the CFO range of \texttt{CSPB.ML.2022} and then train a NN on signals from
\texttt{CSPB.ML.2018} plus frequency-shifted signals from \texttt{CSPB.ML.2018}. This would result in a trained NN that would display
apparent generalization ability when classifying signals from \texttt{CSPB.ML.2022}.

We note, however, that the CFO range corresponding to the signals in the \texttt{CSPB.ML.2022} dataset is non-intersecting with that corresponding
to signals in the \texttt{CSPB.ML.2018} dataset, which allows us to assess a trained NN's ability to generalize what it has learned from dataset
\texttt{CSPB.ML.2018} when it attempts to classify modulated signals from dataset \texttt{CSPB.ML.2022}. If a NN is trained on a large portion
of \texttt{CSPB.ML.2018} and its performance when classifying a remaining subset of \texttt{CSPB.ML.2018} is high, but its performance when
classifying \texttt{CSPB.ML.2022} is low, then that NN's robustness and ability to generalize is low. Likewise, if the classification performance
on both \texttt{CSPB.ML.2018} and \texttt{CSPB.ML.2022} is high, then its generalization ability is also high.

\vspace{-0.1cm}
\section{Network Training and Numerical Results}\label{sec:NumericalResults}
\vspace{-0.1cm}
The proposed capsule network (CAP) and the alternative CNN have been implemented in MATLAB and trained on a high-performance
computing cluster with 18~NVidia V100 graphical processing unit (GPU) nodes available, with each node having 128 GB of memory.
We note that while the NN training process is computationally intensive, if the available computing resources are leveraged appropriately
such that the entire training dataset is loaded into the available memory, training can be completed in several minutes.

\begin{figure}
\begin{center}
{\includegraphics[width=\linewidth]{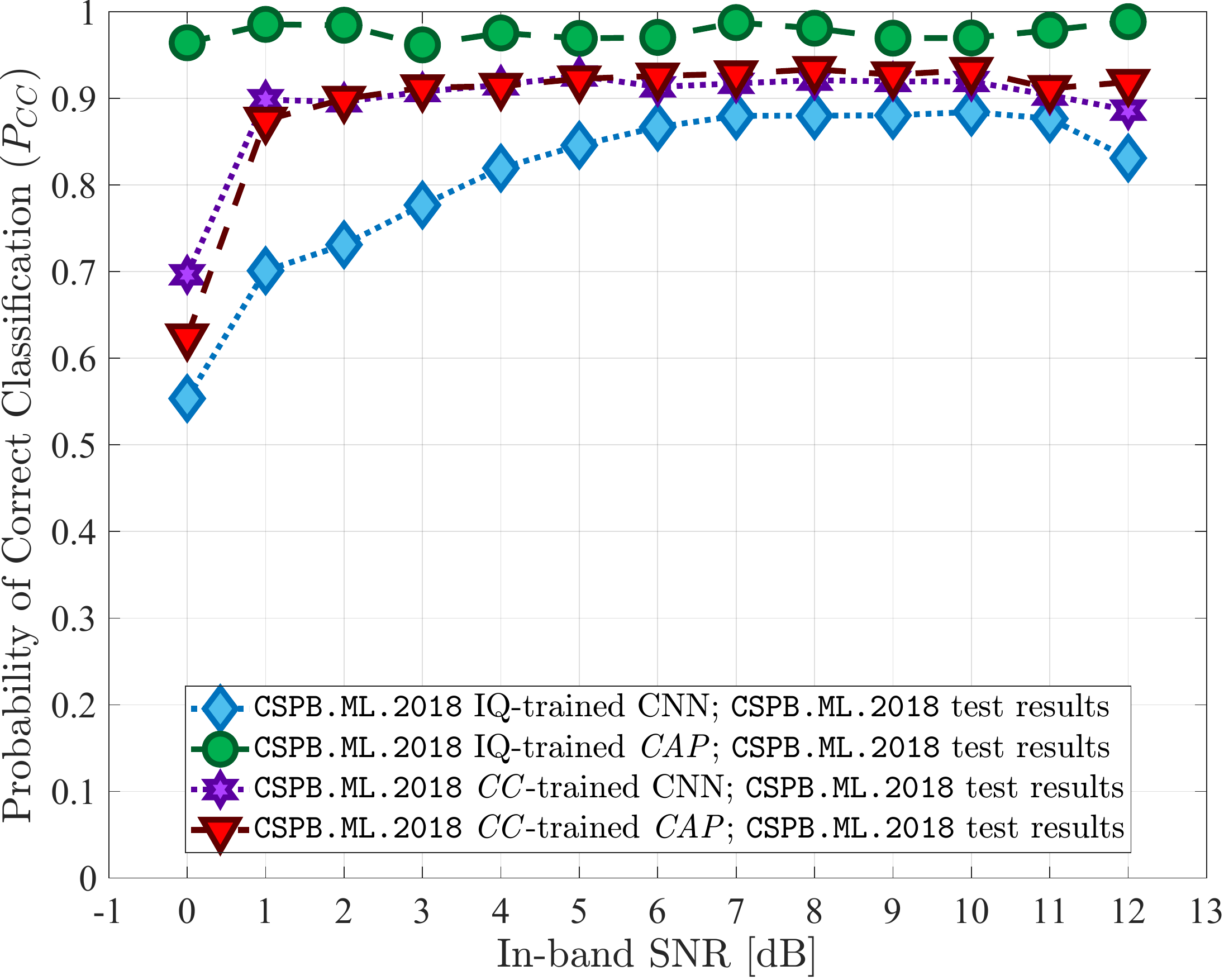}}
\caption{\vspace{-0.25cm}
Performance of NNs trained and tested on \texttt{CSPB.ML.2018} dataset.}\label{fig:SNR_CTest_CTrained_All}
\end{center}
\vspace{-0.25cm}
\end{figure}

\begin{figure}
\begin{center}
{\includegraphics[width=\linewidth]{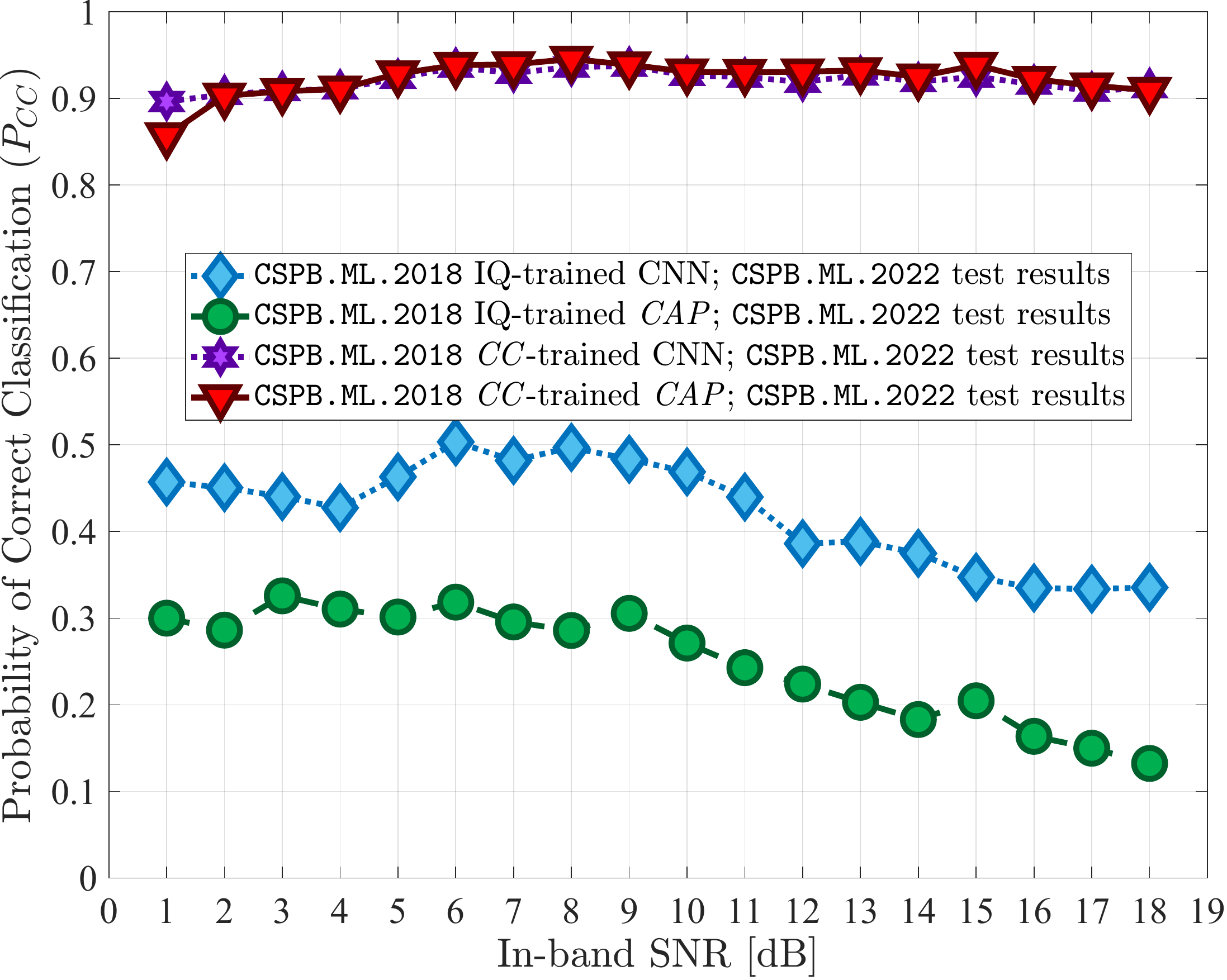}}
\caption{Generalization test results for NNs trained on dataset \texttt{CSPB.ML.2018} and
tested on \texttt{CSPB.ML.2022}.}\label{fig:SNR_GCTest_CTrained_All}
\end{center}
\vspace{-0.75cm}
\end{figure}

\begin{figure}
\begin{center}
{\includegraphics[width=\linewidth]{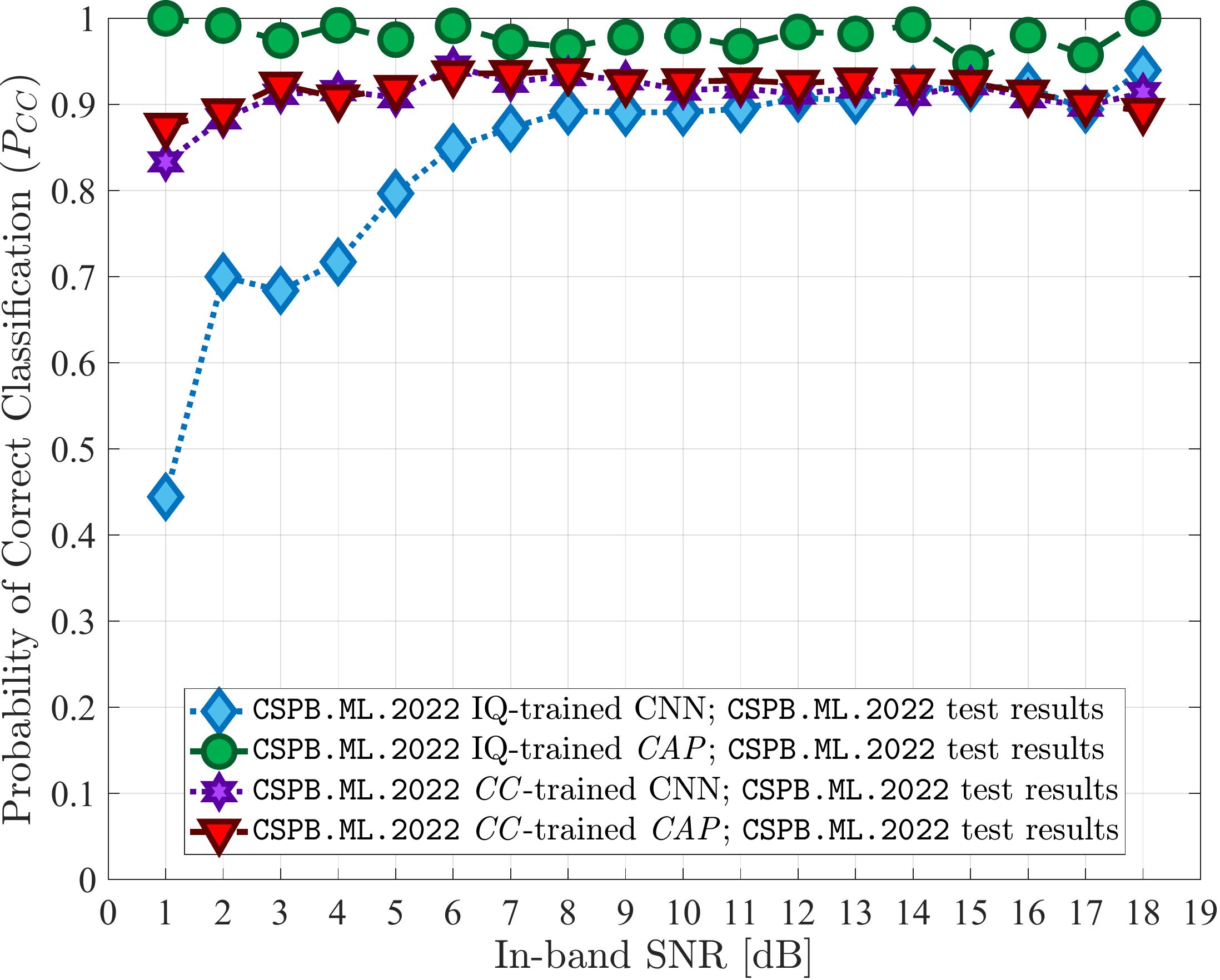}}
\caption{Performance of NNs trained and tested on \texttt{CSPB.ML.2022} dataset.}\label{fig:SNR_GCTest_GCTrained_All}
\end{center}
\vspace{-0.25cm}
\end{figure}

\begin{figure}
\begin{center}
{\includegraphics[width=\linewidth]{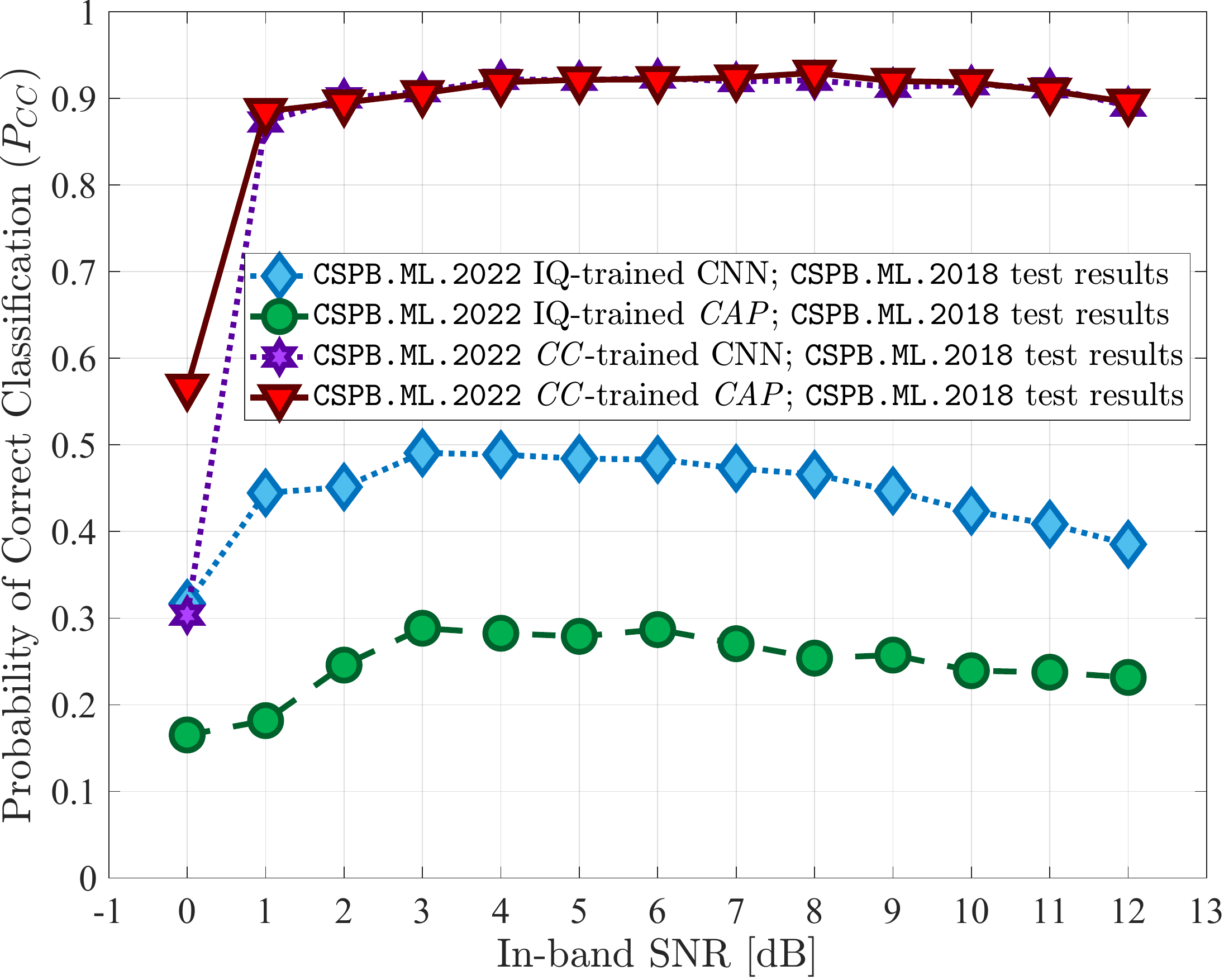}}
\caption{Generalization test results for NNs trained on dataset \texttt{CSPB.ML.2022} and
tested on \texttt{CSPB.ML.2018}.}\label{fig:SNR_CTest_GCTrained_All}
\end{center}
\vspace{-0.5cm}
\end{figure}

Each capsule and convolutional neural network was trained and tested two separate times as follows:
\begin{itemize}
\item In the first training instance, dataset \texttt{CSPB.ML.2018} was used, splitting the available signals into $70\%$ for training,
$5\%$ for validation, and $25\%$ for testing. The probability of correct classification for the test results obtained using the
$25\%$ test portion of signals in \texttt{CSPB.ML.2018} is shown in Fig.~\ref{fig:SNR_CTest_CTrained_All}.
\item The NNs trained on \texttt{CSPB.ML.2018} are then tested on dataset \texttt{CSPB.ML.2022} to assess the generalization abilities of
the trained NN when classifying all signals available in \texttt{CSPB.ML.2022}. The probability of correct classification for this test is shown
in Fig.~\ref{fig:SNR_GCTest_CTrained_All}.
\item In the second training instance, the NN was reset and trained anew using signals in dataset \texttt{CSPB.ML.2022}, with a similar split
of $70\%$ signals used for training, $5\%$ for validation, and $25\%$ for testing. The probability of correct classification for the test results
obtained using the $25\%$ test portion of signals in \texttt{CSPB.ML.2022} is shown in Fig.~\ref{fig:SNR_GCTest_GCTrained_All}.
\item The NNs trained on \texttt{CSPB.ML.2022} are then tested on dataset \texttt{CSPB.ML.2018} to assess the generalization abilities of
the trained NN when classifying all signals available in \texttt{CSPB.ML.2018}. The probability of correct classification for this test is shown
in Fig.~\ref{fig:SNR_CTest_GCTrained_All}.
\end{itemize}


We have also included in Figs.~\ref{fig:SNR_CTest_CTrained_All}-\ref{fig:SNR_CTest_GCTrained_All} the probability of correct
classification corresponding to capsule networks (CAP) and CNNs using the I/Q data for signal classification, which is presented
in our related work \cite{Latshaw_COMM2022, Snoap_CCNC_2022}.

From the plots shown in Fig.~\ref{fig:SNR_CTest_CTrained_All} and Fig.~\ref{fig:SNR_GCTest_GCTrained_All}, we note that all NNs
exhibit high classification performance when trained and tested with signals taken from the same dataset, and that capsule networks
(CAPs) outperform CNNs.

From the plots shown in Fig.~\ref{fig:SNR_GCTest_CTrained_All} and Fig.~\ref{fig:SNR_CTest_GCTrained_All}, we note that NNs
using CC features have robust classification performance and good generalization ability, displaying high classification performance
when trained with signals taken from one dataset and tested on signals in the other dataset. Furthermore, capsule networks continue
to outperform CNNs when CC features are used for signal classification.

By contrast, as already reported in \cite{Latshaw_COMM2022, Snoap_CCNC_2022}, NNs using the I/Q data for classification
of digitally modulated signals do not display generalization abilities and perform poorly when tested with signals that are not in the
same dataset as the ones used for training.


The overall probability of correct classification ($P_{CC}$) for all experiments performed is summarized in Table~\ref{table:OverallPerformance},
providing an alternative perspective on the robustness and generalization performance of NNs that use CC features for classifying digitally
modulated signals, confirming that capsule networks outperform CNNs.

The reason that the CNN and CAP networks generalize so well for the CC input is that the effect of the carrier-frequency offset is
nullified by the blind feature extraction step. That is, the CC features for $x(t) = s(t) e^{j (2\pi f_1 t + \phi_1)} + w(t)$ and
$y(t) = s(t) e^{j (2 \pi f_2 t + \phi_2)} + w(t)$ are identical to within measurement error. So the disjoint probability density functions
for the offsets in the two datasets have little effect on performance. It is much less clear why the networks that are trained on IQ data
do not generalize, or if such networks or their training can be modified to induce both good performance and good generalization.

\vspace{-0.2cm}
\section{Conclusions}\label{sec:Conclusion}
This paper presents a novel ML-based classifier for digitally modulated signals, 
which uses capsule networks and blindly estimated CC features.
The proposed classifier is robust with good generalization abilities and displays 
high classification accuracy when tested with signals coming
from datasets that are different from the training dataset. Future research will 
include a comparison between the performance of the presented
classifier and that of a conventional CSP-based blind classifier.

A key research question suggested by the present work will also be a prime focus:
Why do trained convolutional neural networks (CNNs) that use time-domain signal samples 
(IQ data) as input fail to produce good performance and generalization relative to
CNNs trained on fundamental probabilistic features such as cyclic cumulants? What
is preventing the CNN from extracting these very features, which
we have shown to provide excellent classification performance, thereby
resulting in lower error than the features actually extracted by the CNN for IQ data?
The CNNs are clearly not minimizing the error, since the CNNs that use
cyclic cumulants as input produce smaller classification error. We suspect that
the typical layers used in networks for modulation recognition rely too heavily
on layers borrowed from image-processing applications, and we will seek to modify
the structure of the networks to better match the signal-processing problem at hand.
Obtained insights into this question for CNNs may carry over to the 
IQ-trained capsule networks, which are better-performing but have even lower 
generalization than the CNNs.

\begin{table}[!tp]
\centering
\caption{Classification Performance}
{
\vspace{-0.25cm}
\begin{tabular}{ c c c }
\hline\hline
Classification Model & \begin{tabular}{@{}c@{}} \texttt{CSPB.ML.2018} \\ Test Results\end{tabular} & \begin{tabular}{@{}c@{}} \texttt{CSPB.ML.2022} \\ Test Results\end{tabular} \\
\hline
\begin{tabular}{@{}c@{}} \texttt{CSPB.ML.2018} \\ I/Q-trained CNN\end{tabular} & $85.8\%$ & $41.6\%$ \\
\begin{tabular}{@{}c@{}} \texttt{CSPB.ML.2022} \\ I/Q-trained CNN\end{tabular} & $45.3\%$ & $89.1\%$ \\
\\
\begin{tabular}{@{}c@{}} \texttt{CSPB.ML.2018} \\ I/Q-trained CAP\end{tabular} & $97.5\%$ & $23.7\%$ \\
\begin{tabular}{@{}c@{}} \texttt{CSPB.ML.2022} \\ I/Q-trained CAP\end{tabular} & $25.7\%$ & $97.7\%$ \\
\\
\begin{tabular}{@{}c@{}} \texttt{CSPB.ML.2018} \\ CC-trained CNN\end{tabular} & $91.4\%$ & $92.4\%$ \\
\begin{tabular}{@{}c@{}} \texttt{CSPB.ML.2022} \\ CC-trained CNN\end{tabular} & $91.3\%$ & $91.8\%$ \\
\\
\begin{tabular}{@{}c@{}} \texttt{CSPB.ML.2018} \\ CC-trained CAP\end{tabular} & $92.3\%$ & $93.1\%$ \\
\begin{tabular}{@{}c@{}} \texttt{CSPB.ML.2022} \\ CC-trained CAP\end{tabular} & $91.6\%$ & $92.5\%$ \\
\hline\hline
\end{tabular}
}\label{table:OverallPerformance}
\vspace{-0.3cm}
\end{table}

\section*{Acknowledgment}
The authors would like to acknowledge the use of Old Dominion University High-Performance Computing facilities for obtaining numerical results presented in this work.

\IEEEtriggeratref{13}


 \bibliographystyle{IEEEtran}
 \bibliography{milcom2022_arXiv}
%
%
%

\end{document}